
\magnification = \magstep1
\baselineskip = 22pt plus 2pt minus 1pt
\hsize 14truecm

\font\fontA=cmss10 scaled\magstep3
\font\fontB=cmr10
\font\fontC=cmssi10
\font\fontE=cmcsc10

\fontB

\centerline{\fontA Density Matrix Renormalization Group }
\centerline{\fontA Method for 2D Classical Models }
\vskip 15pt
\centerline{Tomotoshi {\fontE Nishino}}
\centerline{\it Department of Physics, Faculty of
Science,}
\centerline{\it Tohoku University, Sendai 980-77}
\vskip 20pt
\centerline{(~~~~~~~~~~~~~~~~~~~~~~~~~~~~~~~~~~~~~~~~~)}
\vskip 10pt
\centerline{\fontC Synopsis}
\vskip  5pt


{\narrower
The density matrix renormalization group (DMRG) method is
applied to the interaction round a face (IRF) model. When
the transfer matrix is asymmetric, singular-value
decomposition of the density matrix is required. A trial
numerical calculation is performed on the square lattice
Ising model, which is a special case of the IRF model.  }
\vskip 1truecm

\leftline{ Submitted 12 March, 1995 }
\leftline{   Reviced 19  July, 1995 }
\leftline{  to be appear in J. Phys. Soc. Jpn. Vol.64,
No.10, 1995.}
\vskip 1truecm

\item{e-mail:} nishino@kaws.coge.tohoku.ac.jp

\vfill\eject

The renormalization group is one of the basic concepts in
statistical mechanics.$^{1,2)}$ Its real-space expression
--- the real-space renormalization group$^{3)}$ --- has
been applied to various statistical models.  Recently White
proposed a precise numerical renormalization algorithm,
which is known as the density matrix renormalization group
(DMRG) method.$^{4,5)}$ This method has been widely used
for one-dimensional (1D) quantum lattice models$^{5-7)}$
because of its numerical accuracy and portability.

The DMRG method was originally designed for 1D quantum
lattice models. Since $d$-dimensional quantum systems are
closely related to classical systems in $d$+$1$
dimensions,$^{8)}$ the DMRG method is applicable to
two-dimensional (2D) classical systems. In this paper we
apply the DMRG method to the interaction round a face (IRF)
model,$^{9)}$ which contains various 2D statistical models
such as the Ising model and the vertex models.

The IRF model is defined by a Boltzmann weight $W$ on each
face, which is surrounded by four $n$-state spins $\sigma$.
The transfer matrix of the IRF model is given by
$$
T^{(2N)}(\sigma'_1\ldots\sigma'_{2N}
| \sigma_1^{}\ldots\sigma_{2N}^{}) =
\prod_{i=1}^{2N-1}
W(\sigma'_i\sigma'_{i+1} | \sigma_i^{}\sigma_{i+1}^{}) ,
\eqno(1)
$$
where $\sigma_i^{}$ and $\sigma_i^{'}$ are $n$-state spins
on adjacent rows with width $2N$. (Fig.1({\bf a})) We
assume open boundary conditions. In general, the transfer
matrix is not symmetric. If the Boltzmann weights satisfy a
constraint --- the Yang-Baxter relation$^{9,10)}$ --- the
model is analytically solvable. Here, we consider a
numerical analysis of $T^{(2N)}$ in view of the existence
of unsolvable cases.

The DMRG method for the IRF model is expressed as a
renormalization of the transfer matrix $T^{(2N)}$. Figures
1({\bf a}) and 1({\bf b}) show the method of
renormalization. First of all, we assume that $T^{(2N)}$
can be renormalized into a product form
$$
\eqalignno{
&{\tilde T}^{(2N)}(\xi'_L \sigma'_L \sigma'_R \xi'_R |
\xi_L^{} \sigma_L^{} \sigma_R^{} \xi_R^{}) \cr
&= {\tilde T}^{(N)}_L(\xi'_L \sigma'_L |
\xi_L^{} \sigma_L^{})
W(\sigma'_L \sigma'_R | \sigma_L^{} \sigma_R^{})
{\tilde T}^{(N)}_R(\sigma'_R \xi'_R | \sigma_R^{} \xi_R^{})
, &(2)}
$$
where ${\tilde T}^{(N)}_L$ and ${\tilde T}^{(N)}_R$
represent renormalized transfer matrices for the left and
the right half of the system, respectively. (Fig.1({\bf
a})) The block-spin variables  $\xi_L^{}$ and $\xi_R^{}$
correspond to groups of the $n$-state spins
$\{\sigma_1^{}\ldots\sigma_{N-1}^{}\}$ and
$\{\sigma_{N+1}^{}\ldots\sigma_{2N}^{}\}$, respectively. We
assume that $\xi_L^{}$ and $\xi_R^{}$ take $m$ different
states, where $m$ is much smaller than $n^{N-1}$. If there
is a mapping from ${\tilde T}^{(2M)}$ to ${\tilde
T}^{(2M+2)}$ for arbitrary $M$, then ${\tilde T}^{(2N)}$ in
eq.(2) is obtained through successive mapping:
$$
T^{(4)} = {\tilde T}^{(4)}
\rightarrow {\tilde T}^{(6)} \ldots
{\tilde T}^{(2N-2)} \rightarrow {\tilde T}^{(2N)}.
\eqno(3)
$$
In each step, block-spin transformations
$\{\xi_L^{}\sigma_L^{}\} \rightarrow \xi_L^{\rm new}$ and
$\{\sigma_R^{} \xi_R^{}\} \rightarrow \xi_R^{\rm new}$ are
required.  (Fig.1({\bf b})) The block-spin transformations
are given in the following. Since the argument is common to
both $\xi_L^{}$ and $\xi_R^{}$, we discuss the
renormalization for $\xi_L^{}$ only.

The matrix dimension of ${\tilde T}^{(2M)}$ is $(nm)^2$
for arbitrary $M$. If $m$ is small enough, we can
numerically solve the eigenvalue problem
$$
\sum_{\xi_L^{} \sigma_L^{} \sigma_R^{} \xi_R^{}}
{\tilde T}^{(2M)}(\xi'_L \sigma'_L \sigma'_R \xi'_R |
\xi_L^{} \sigma_L^{} \sigma_R^{} \xi_R^{})
\Psi(\xi_L^{} \sigma_L^{} \sigma_R^{} \xi_R^{})
= {\tilde \lambda}^{(2M)}
\Psi(\xi'_L \sigma'_L \sigma'_R \xi'_R) ,
\eqno(4)
$$
and obtain the `right' eigenvector $\Psi$ that corresponds
to the largest eigenvalue ${\tilde \lambda}^{(2M)}$. In the
same way, we obtain the `left' eigenvector $\Phi$ that
satisfies $\Phi{\tilde T}^{(2M)} = \Phi {\tilde
\lambda}^{(2M)}$. Note that if  ${\tilde T}^{(2M)}$ is
asymmetric, $\Phi$ is not the complex conjugate of $\Psi$.
The density matrix$^{4,5)}$ is then expressed as a partial
product between $\Psi$ and $\Phi$:
$$
\rho_L^{}(\xi'_L \sigma'_L | \xi_L^{} \sigma_L^{})
= \sum_{\sigma''_R \xi''_R}
\Phi (\xi'_L \sigma'_L \sigma''_R \xi''_R)
\Psi (\xi_L^{} \sigma_L^{} \sigma''_R \xi''_R) .
\eqno(5)
$$
Apart from the density matrix for 1D quantum systems,
$\rho_L$ is not always Hermitian. Therefore we must perform
the singular-value decomposition of $\rho_L$
$$
\sum_{\xi'_L \sigma'_L \xi_L^{} \sigma_L^{}}
O( i | \xi'_L \sigma'_L)
\rho_L^{}(\xi'_L \sigma'_L | \xi_L^{} \sigma_L^{})
Q( \xi_L^{} \sigma_L^{} | j )
= \delta_{ij}^{} \omega_j^{} ,
\eqno(6)
$$
and obtain $O$ and $Q$ that satisfy
$
\sum_{\xi_L^{} \sigma_L^{}} O( i | \xi_L \sigma_L)
Q( \xi_L^{} \sigma_L^{} | j ) = \delta_{ij}^{}
$.
We assume the decreasing order of the singular values:
$\omega_1^{} \ge \omega_2^{} \ge \ldots \omega_{nm}^{}$.
The $m$ by $nm$ matrix  $O( \xi_L^{\rm new} | \xi_L^{}
\sigma_L^{} )$ together with the the $nm$ by $m$ one $Q(
\xi_L^{} \sigma_L^{} | \xi_L^{\rm new} )$ represent the
block-spin transformation from $\xi_L^{}\sigma_L^{}$ to a
new $m$-state block-spin $\xi_L^{\rm new}$. The
transformation naturally gives a relation between ${\tilde
T}^{(M)}_L$ and ${\tilde T}^{(M+1)}_L$, which is a linear
transformation
$$
\eqalignno{
&{\tilde T}^{(M+1)}_L({\xi'}_L^{\rm new} \sigma'_L |
\xi_L^{\rm new} \sigma_L^{}) = \cr
&\sum_{\xi'_L \sigma' \xi_L^{} \sigma}
O( {\xi'}_L^{\rm new} | \xi'_L \sigma' )
{\tilde T}^{(M)}_L(\xi'_L \sigma' | \xi_L^{} \sigma )
W( \sigma' \sigma'_L | \sigma \sigma_L^{} )
Q( \xi_L^{} \sigma | \xi_L^{\rm new} ) .
&(7)}
$$
We also obtain a similar relation between ${\tilde
T}^{(M)}_R$ and ${\tilde T}^{(M+1)}_R$ in the same way. Now
we obtain the renormalization processes in eq.(3).

As one  repeats the mapping in eq.(3) with the aid of
eqs.(4)-(7), the eigenvectors of the renormalized transfer
matrix ${\tilde T}^{(2M)}$ approach their fixed-point
values. After $\Phi$ and $\Psi$ have converged, we can
obtain local thermodynamic quantities in the large-$N$
limit. For example,  the nearest-neighbor spin correlation
is expressed as
$$
\langle \sigma_i^{} \sigma_{i+1}^{} \rangle \sim
\langle \sigma_L^{} \sigma_R^{} \rangle
= \sum_{\xi_L^{} \sigma_L^{} \sigma_R^{} \xi_R^{}}
 \Phi(\xi_L^{} \sigma_L^{} \sigma_R^{} \xi_R^{})
\sigma_L^{} \sigma_R^{}
\Psi(\xi_L^{} \sigma_L^{} \sigma_R^{} \xi_R^{})
\eqno(8)
$$
when the inner product $(\Phi,\Psi)$ is unity. In the same
way, we can obtain the internal energy, susceptibilities,
and specific heat. Observations of correlation functions
and surface tensions are possible with the use of the
`finite chain algorithm' of the DMRG method.$^{5)}$

We examine efficiency of the DMRG method. As a reference,
we calculate specific heat of the nearest-neighbor Ising
model, where the Boltzmann weight is
$$
W(\sigma'_i\sigma'_{i+1} | \sigma_i^{}\sigma_{i+1}^{})
= {\rm exp}\left\{ {K \over 2}( \sigma'_i\sigma'_{i+1}
+ \sigma'_i\sigma_i^{} +
\sigma'_{i+1}\sigma_{i+1}^{}
+ \sigma_i^{}\sigma_{i+1}^{}) \right\}
\eqno(9)
$$
for $\sigma = \pm 1$. In this case, the renormalized
transfer matrix ${\tilde T}^{(2N)}$ is real-symmetric for
arbitrary $N$, and therefore the eigenvectors $\Phi$ and
$\Psi$ in eq.(5) are the same. The specific heat ${\rm
C_v}$ is obtained from a numerical differential of the
nearest-neighbor spin correlation $\langle \sigma_i^{}
\sigma_{i+1}^{} \rangle$ in eq.(8). The numerical error
$\epsilon$ in ${\rm C_v}$  is obtained through a comparison
between calculated data and the exact value.$^{11)}$

Figure 2 shows the calculated specific heat. The plotted
data are obtained within 1024 iterations ($2N = 2048$) when
$m = 60$. The numerical error $\epsilon$ in ${\rm C_v}$ is
non-negligible near $T_c$, partly because the numerical
convergence of the bond energy $\langle \sigma_i^{}
\sigma_{i+1}^{} \rangle$ with respect to $N$ and $m$
becomes worse as the parameter $K = J / k_B T$ approaches
its critical value $K^* = J / k_B T_c$. The calculated
energy per site at $T = T_c$ is $-1.41398$ when $m = 60$,
which is comparable in accuracy to a recent numerical
estimate$^{12)}$ $-1.419(1)$ by a microcanonical Monte
Carlo simulation; the exact value$^{11)}$ is $-\sqrt{2} =
-1.414213$. Another source of error is in the numerical
differential of $\langle \sigma_i^{} \sigma_{i+1}^{}
\rangle$ with respect to $T$. Since the derivative diverges
at $T_c$, a slight error in $\langle \sigma_i^{}
\sigma_{i+1}^{} \rangle$ reduces the accuracy in ${\rm
C_v}$ near $T_c$.

The numerical result shows that the DMRG method is quite
efficient in the off-critical --- high- and low-temperature
--- regions. On the other hand, it is difficult to maintain
good numerical accuracy near the critical point, because
long-range spin correlations prevent us from obtaining good
renormalized transfer matrix, whose size $(=2m)$ is limited
by computational restrictions. This convergence property of
the DMRG method for the Ising model is in accordance with
the fact that the DMRG method works more efficiently for 1D
quantum systems with a finite excitation gap.$^{5)}$

I would like  to thank S.~R.~White and M.~Guerreo for
helpful comments and discussions. I also thank Y.~Akutsu,
M.~Kikuchi, K.~Okunishi, and H.~Otsuka for valuable
suggestions. Most of the numerical calculations were done
by using the super computer NEC SX-3/14R at the computer
center of Osaka University.

\vfill\eject

\beginsection{}References

\item{1)} L.~P.~Kadanoff: Physics {\bf 2} (1965) 263.
\item{2)} K.~G.~Wilson and J.~Kogut: Phys. Rep. {\bf 12C}
(1974) 75.
\item{3)} T.W.Burkhardt and J.M.J. van Leeuwen:
{\it Real-Space Renormalization,} Topics in Current
Physics vol.{\bf 30}, (Springer, Berlin, 1982).
\item{4)} S.~R.~White: Phys. Rev. Lett. {\bf 69} (1992)
2863.
\item{5)} S.~R.~White: Phys. Rev. {\bf B48} (1993) 10345.
\item{6)} C.~C.~Yu and S.~R.~White: Phys. Rev. Lett.
{\bf 71} (1993) 3866.
\item{7)} R.~M.~Noack, S.~R.~White and D.~J.~Scalapino:
Phys. Rev. Lett {\bf 73} (1994) 882.
\item{8)} H.~F.~Trotter: Proc. Am. Math. Soc. {\bf 10}
(1959) 545.
\item{9)} R.~J.~Baxter: {\it Exactly Solved Models in
Statistical Mechanics,} (Academic Press, London, 1980)
p.363.
\item{10)} C.~N.~Yang: Phys. Rev. Lett {\bf 19} (1967)
1312.
\item{11)} L.~Onsager: Phys. Rev. {\bf 65} (1944) 117.
\item{12)} A.~J.~F.~de Souza and F.~G.~Bardy Moreira:
Phys. Rev. {\bf B48} (1993) 9586.

\beginsection{~} Figure Captions

\item{Figs.~1.}
Renormalization processes.  ({\bf a}) We assume that the
transfer matrix $T^{(2N)}$  can be renormalized as the
product form ${\tilde T}^{(2N)}$ in eq.(2). ({\bf b}) The
assumption is justified through the relation between
${\tilde T}_{L}^{(M)}$ and ${\tilde T}_{L}^{(M+1)}$ in
eq.(7).

\item{Fig.~2.}
Numerically calculated specific heat of the Ising model.
The plotted data are obtained within 1024 iterations
($2N=2048$) under the condition $m = 60$. The numerical
error $\epsilon$ in each data point is indicated by the
symbols.

\vfill\bye